# OPINION MINING OF MOVIE REVIEWS AT DOCUMENT LEVEL

Richa Sharma[1], Shweta Nigam[2] and Rekha Jain[3]

[1,2]M.Tech Scholar, Banasthali Vidyapith, Rajasthan, India
[3]Assistant Professor, Banasthali Vidyapith, Rajasthan, India

## ABSTRACT

*The whole world is changed rapidly and using the current technologies Internet becomes an essential need for everyone. Web is used in every field. Most of the people use web for a common purpose like online shopping, chatting etc. During an online shopping large number of reviews/opinions are given by the users that reflect whether the product is good or bad. These reviews need to be explored, analyse and organized for better decision making. Opinion Mining is a natural language processing task that deals with finding orientation of opinion in a piece of text with respect to a topic. In this paper a document based opinion mining system is proposed that classify the documents as positive, negative and neutral. Negation is also handled in the proposed system. Experimental results using reviews of movies show the effectiveness of the system.*

## 1. INTRODUCTION

Large number of user reviews or suggestions on everything is present on the web nowadays; reviews may contain the reviews on products, user or critic reviews on movies etc. which helps other users in their decision making. Reviews are increasing in a faster rate day by day because every person likes to give their opinion on the Web. Large numbers of reviews are available for a single product which makes difficult for a customer to read all the reviews and make a decision. Thus, mining this data, identifying the user opinions and classify them is an important task. Opinion Mining is a *Natural Language Processing (NLP)* and *Information Extraction (IE)* task that aims to obtain feelings of the writer expressed in positive or negative comments by analyzing a large number of documents [13]. It combines the techniques of computational linguistics and Information Retrieval (IR). The main task of Sentiment analysis is to classify the documents and determine its polarity. Polarity is expressed as positive, negative or neutral. There are three levels on which sentiment analysis can be performed: [4]

- Document level: Classifies the whole document as positive, negative or neutral and commonly known as document-level sentiment classification.
- Sentence level: Classifies the sentences as positive, negative or neutral commonly known as sentence-level sentiment classification.
- Aspect & Feature level: Classifies sentences/documents as positive, negative or neutral based on the aspects of those sentences/documents commonly known as aspect-level sentiment classification.

In this paper an Opinion Mining System is proposed named as "Document based Sentiment Orientation System" based on unsupervised approach that determine the sentiment orientation of documents. Sentiment orientation determines the polarity of documents, it classifies the documents as positive and negative [3][14]. This approach helps the users in decision making by providing the summary of total number of positive and negative documents. Proposed approach extracts the opinion words from the documents and determines the corresponding polarity of the documents. Figure 1 presents an example of document based opinion mining.

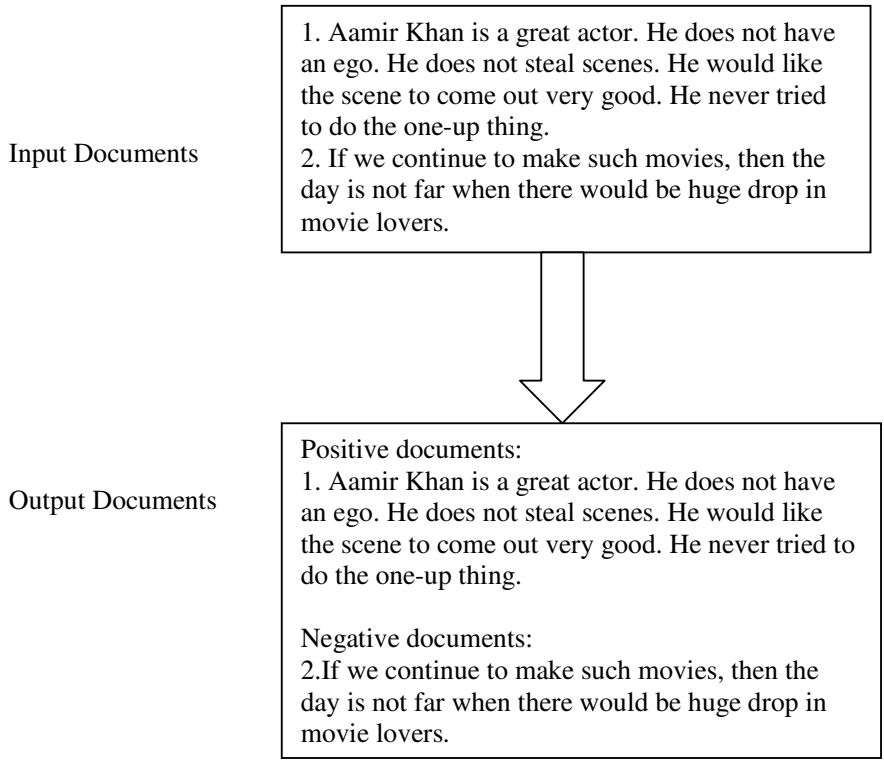

Figure 1 Example of Document based Opinion Mining

Negation is also handled in this approach. WordNet is used as a dictionary to determine the synonyms and antonyms of opinion words. The rest of the paper is organized as follows: Section 2 discusses related work. Section 3 describes the proposed approach. Section 4 shows the experimental results of the system. Section 5 concludes the paper.

## 2. EXISTING RESEARCH WORK

Existing researches in the document based opinion mining are mentioned below. The most prominent work was done by Turney [12]. "Poor" and "Excellent" seed words are used by him to calculate the semantic orientation, point wise mutual information method is used to calculate the semantic orientation. The sentiment orientation of a document was calculated as the average semantic orientation of all such phrases. 66% accuracy was achieved for the movie review domain.

Harb et al. [1] used two sets of seed words with positive and negative semantic orientations to perform blog classification .Google's search engine is used to create association rules. Total number of positive and negative adjectives is counted in a document to classify the documents.

They achieved 0.717 F1 score identifying positive documents and 0.622 F1 score identifying negative documents.

Taboada et al. [10] used lexicon-based method to perform sentiment classification. For classification positive and negative words dictionaries are used and semantic orientation calculator (SO-CAL) is built that incorporate intensifiers and negation words. This approach has been shown to have 59.6% to 76.4% accuracy on 1900 documents of the movie review dataset. Andrea Esuli et al. [2] proposed semi-supervised learning method started from expanding an initial seed set using WordNet. Semantic orientation is determined through gloss classification by statistical technique.

Zagibalov et al. [19] perform unsupervised sentiment classification of product reviews in Chinese using automatic seed word selection method. Method requires information about commonly occurring negations and adverbials in order to iteratively and sentiment bearing items. The results obtained are close to those of supervised classifiers and sometimes better, up to an F1 score of 92%.

Ting-Chun Peng et al.[16] extracts sentiment phrases of each review by using part of speech patterns, as a query term they used unknown sentiment phrase and retrieve top-N relevant phrases from a search engine. Then, based on the sentiments of nearby known relevant phrase using lexicons, sentiments of unknown sentiment phrases are computed

Chunxu Wu [7] proposed an approach to determine the orientation of opinion by using semantic similarity measures. Semantic Orientation of context independent opinions is determined and the context dependent opinions using linguistic rules to infer orientation of context distinct-dependent opinion is considered. Contextual information from other reviews that comment on the same product feature to determine the context indistinct-dependent opinions were extracted.

## 3. PROPOSED SYSTEM

The unsupervised dictionary based technique is used in this system. WordNet is used as a dictionary to determine the opinion words and their synonyms and antonyms [8]. The proposed work is closely related to the Minqing Hu and Bing Liu work on Mining and Summarizing Customer Reviews [11]. Figure 2. gives the overview of the proposed system 'Document based Sentiment Orientation System'[15]. User and critic reviews of the movies were collected and applied as an input to the system. The system classifies each document as positive, negative and neutral and presents the total number of positive, negative and neutral number of documents separately in the output. The output generated by the system helpful for the users in decision making, they can easily identify how many positive and negative documents are present. The polarity of the given documents is determined on the basis of the majority of opinion words.

Proposed system is divided into following phases:

### 3.1.Data Collection

Large numbers of movie reviews are collected from different-different websites. Movie reviews contain the user and critic reviews, there are various websites available on the web which contain movie reviews like movies.ndtv.com [12], www.rottentomatoes.com [17], www.imdb.com[18] etc. Before determining the polarity of the collected reviews, pre-processing of the collected reviews are necessary to get the cleaned reviews. Pre-processed reviews are applied as input.

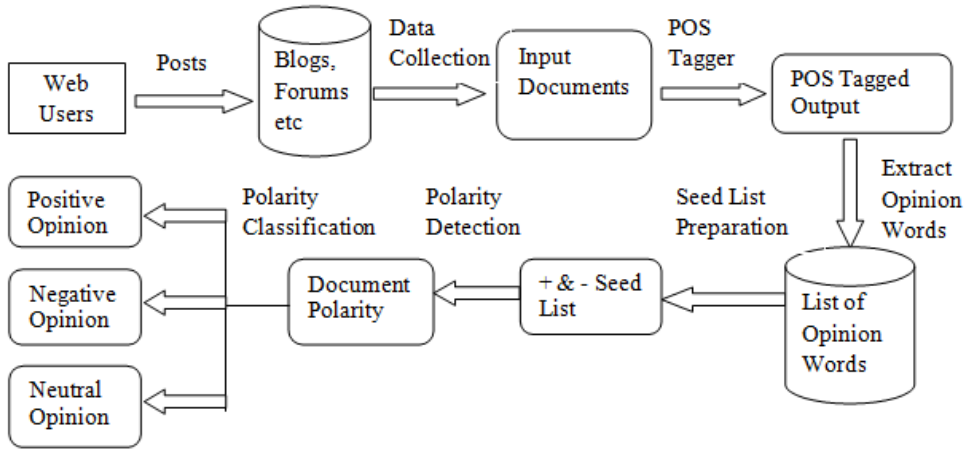

Figure 2 Document based Sentiment Orientation System

### 3.2. POS Tagging

Collected reviews are sent to the POS tagger that tags all the words of the documents to their appropriate part of speech tag [6][9]. POS tagging is necessary to determine the opinion words. It can be done manually or with the help of POS tagger. POS tagger is used here to tag the entire document. For Example Figure 3 shows an example of POS tagging.

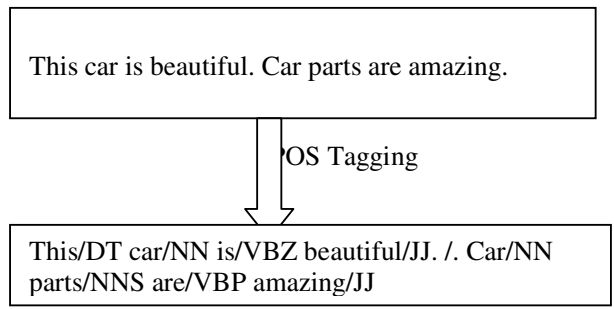

Figure 3 Example of POS Tagging

### 3.3. Extracting Opinion Words and Seed List Preparation

Seed list initially contains some of the opinion words along with their polarity. From the tagged output all the opinion words were extracted. The extracted opinion words matched with the words stored in seed list .If the word is not found in the seed list then the synonyms are determined with the help of WordNet. Each synonym is matched with words in the seed list, if any synonym matched then extract opinion word is stored with the same polarity in the seed list. If none of the synonym is matched then the antonym is determined from the WordNet and the same process are repeated, if any antonym matched then extract opinion word is stored with the opposite polarity in the seed list. In this way the seed list keeps on increasing. It grows every time whenever the synonyms or antonyms words are found in WordNet matches with seed list.

**3.4. Polarity Detection and Classification**

With the help of seed list and Wordnet, the polarity of the documents is determined. Polarity is determined on the basis of majority of opinion words, if the number of positive words is more, then the polarity of document is positive otherwise the polarity is negative and if the number of positive and negative words is equal then the document shows the neutral polarity. The system classifies the document into one of the three categories as:

- **Positive Opinion** – The system gives the positive opinion to the document if the number of positive opinion words is more than the number of negative opinion words in the document. For example document like

    "The story of the movie is **good** but the acting of the actors is *awful*. Songs of the movie **hit** the chartbusters, youngsters **likes** the songs very much."

    This document shows positive polarity because here the positive opinion shown in bold is greater than the negative opinion words shown in italic bold.

- **Negative Opinion** – The system gives the negative opinion to the document if the number of negative opinion words is more than the number of positive opinion words in the document. For example document like

    "In this year mostly movies are *flop* because now the viewers get *bored* of watching these kinds of predictable stories. There are only few **good** movies have come this year."

    This document shows negative polarity because here the negative opinion shown in italic bold is greater than the positive opinion words shown in bold.

- **Neutral Opinion** – The system gives the neutral opinion to the document if the number of positive opinion words is equal to the number of negative opinion words in the document.

    "I **like** commercial movies but I get *bored* easily. I watch movies when I am free"

    This document shows neutral polarity because here the negative opinion shown in italic bold is equal to the positive opinion words shown in bold.

As negation is also handled in this system, so if the opinion word is preceded by not then the polarity of review is reversed. For example the sentence

   "This movie is not good." shows negative polarity ,here good is a positive opinion word but it is preceded by not so the polarity of the sentence is reversed.

## 4. EXPERIMENTAL RESULTS

Critics and user reviews of the movies were used to perform the experiment. All the collected reviews applied to the proposed system which classifies the reviews as positive, negative and neutral. Final results are presented in graphical charts. To compute how well the system classifies each document as compared to human decision, all the documents were manually classified and the corresponding opinion was determined. The results were then compared with the results of the system. Same reviews were also applied to the other system named as "AIRC Sentiment Analyzer" available online. Finally the results of the two systems were compared and the results have shown that the results of the Document based Sentiment Orientation System are better than that of AIRC Sentiment Analyzer. Three evaluation measures are used on the basis of which systems are compared, these are:-

- Precision
- Recall
- Accuracy

Table 1 gives the precision, recall and accuracy results of the current system.
Table 2 gives the precision, recall and accuracy results of the system 'AIRC Sentiment Analyzer'
Table 3 presents the comparison of the two systems which shows that performance of 'Document based Sentiment Orientation System' is better than the 'AIRC Sentiment Analyzer
Figure 4 presents the precision, accuracy and recall results of current system in graphical form.
Figure 5 presents the precision, accuracy and recall results of 'AIRC Sentiment Analyzer' in graphical form.
Figure 6 represents the comparison of the results of the two systems which shows that performance of 'Document based Sentiment Orientation System' is better than the 'AIRC Sentiment Analyzer' in graphical form.

Table 1. 'Document based Sentiment Orientation System' Results

| **Measures** | **Results** |
| --- | --- |
| Accuracy | 0.63 |
| Precision | 0.63 |
| Recall | 0.7 |

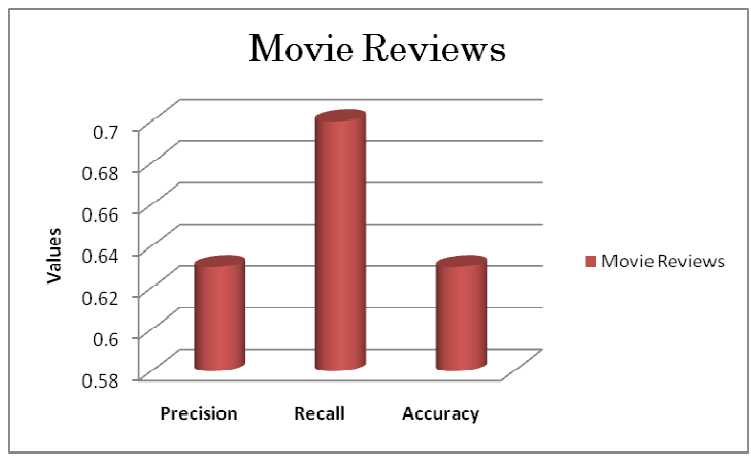

Figure 4 'Document based Sentiment Orientation System' Graph

Table 2. 'AIRC Sentiment Analyzer' Results

| Measures | Results |
|---|---|
| Accuracy | 0.58 |
| Precision | 0.52 |
| Recall | 0.6 |

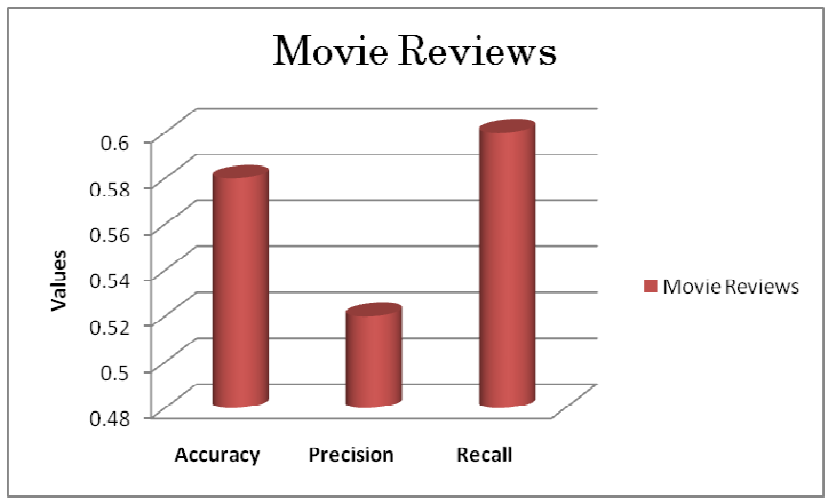

Figure 5 'AIRC Sentiment Analyzer' Graph

Table 3. Comparison of Document based Sentiment Orientation System & AIRCS

| Measures\System | AIRC | Sentiment Orientation System |
|---|---|---|
| Accuracy | 0.58 | 0.63 |
| Precision | 0.52 | 0.63 |
| Recall | 0.6 | 0.7 |

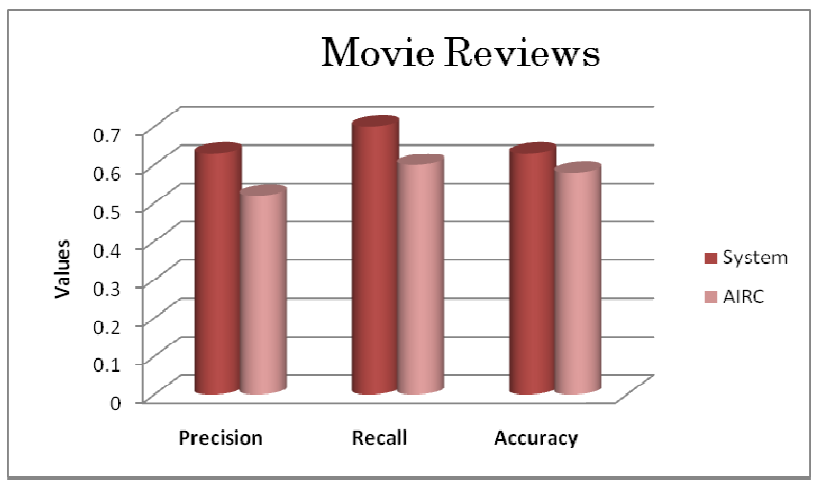

Figure 6. Comparison of Document based Sentiment Orientation System & AIRC Sentiment Analyzer

The above results shows that the 'Document based Sentiment Orientation System' performs well with respect to the movie domain as compared to 'AIRC Sentiment Analyzer'. Proposed system achieved the accuracy of 63%.

## 5. CONCLUSION

The objective of this paper is to determine the polarity of the movie reviews at document level. The results generated by the system are summarized and helpful for the user in decision making. Experimental results indicate that the 'Document based Sentiment Orientation System' perform well in this domain. Opinion mining is very important nowadays from the common man to a businessman,everyone is dependent on the Web. The opinions expressed on the web helps the users to determine which product or movie is good for them and it helps the businessman to determine what the customers thinks about their products. So, it is necessary to mine this large number of reviews and classify them, so it is helpful for them to read and take decisions. In future work, efforts would be done to improve this technique so that it would deal with the documents contain relative clauses like not only-but also , neither-nor, either-or etc.